\documentstyle[procsla,12pt]{article}
\def\Journal#1#2#3#4{{#1} {\bf #2} (#3) #4}

\def\NPB{{\em Nucl.\,Phys.}}
\def\NPBP{{\em Nucl.\,Phys.}\,B (Proc.\,Suppl.)}
\def\PLB{{\em Phys.\,Lett.}}

\def\PRD{{\em Phys.\,Rev.}}
\def\ZPC{{\em Z.\,Phys.}}
\def\bea{\begin{eqnarray}}
\def\eea{\end{eqnarray}}
\def\Bto#1{$B\rightarrow#1\ell\overline{\nu}_\ell$}
\def\Btot#1#2
{B\rightarrow{#1}\ell\overline{\nu}_\ell,\,{#2}\ell\overline{\nu}_\ell}
\def\w{\omega}

\begin{document}

\newcommand{\mev}{\mbox{\rm MeV}}
\newcommand{\gev}{\mbox{\rm GeV}}
\newcommand{\fm}{\mbox{\rm fm}}
\newcommand{\be}{\begin{equation}}
\newcommand{\ee}{\end{equation}}
\newcommand{\bbar}{\mbox{$B^0 - \overline{B^0}${}}}
\newcommand{\plus}{\makebox[15pt][c]{$+$}}
\newcommand{\minus}{\makebox[15pt][c]{$-$}}

\newcommand{\errrr}[2]{\raisebox{0.08em}{\scriptsize {$\;\begin{array}{@{}l@{}}
                          \plus\makebox[1.2em][r]{#1} \\[-0.12em] 
                          \minus\makebox[1.2em][r]{#2} 
                        \end{array}$}}}
\newcommand{\errr}[2]{\raisebox{0.08em}{\scriptsize {$\;\begin{array}{@{}l@{}}
                          \plus\makebox[0.9em][r]{#1} \\[-0.12em] 
                          \minus\makebox[0.9em][r]{#2} 
                        \end{array}$}}}
\newcommand{\err}[2]{\raisebox{0.08em}{\scriptsize {$\;\begin{array}{@{}l@{}}
                          \plus\makebox[0.55em][r]{#1} \\[-0.12em] 
                          \minus\makebox[0.55em][r]{#2} 
                        \end{array}$}}}
\newcommand{\er}[2]{\raisebox{0.08em}{\scriptsize {$\;\begin{array}{@{}l@{}}
                          \plus\makebox[0.15em][r]{#1} \\[-0.12em] 
                          \minus\makebox[0.15em][r]{#2} 
                        \end{array}$}}}

\begin{center}

  {\bf STATUS AND PROSPECTS FOR LATTICE CALCULATIONS\\
       IN HEAVY QUARK PHYSICS\footnote{Invited talk presented at the
         III German-Russian Workshop on
         Heavy Quark Physics, Dubna, Russia, 20-22 May 1996.}\\}

\vspace*{1cm}
HARTMUT WITTIG\\
{\it DESY-IfH, Platanenallee 6, D-15738 Zeuthen, and\\
     HLRZ, c/o Forschungszentrum J\"ulich, D-52425 J\"ulich, Germany}
\end{center}

\vspace*{0.85cm}
\begin{abstracts}
  {\small The current status of lattice calculation of weak matrix
    elements for heavy quark systems is reviewed. After an assessment
    of systematic errors in present simulations, results for the $B$
    meson decay constant, the $B$~parameter $B_B$ and semi-leptonic
    heavy-to-light and heavy-to-heavy transitions are discussed. The
    final topic are lattice results for heavy baryon spectroscopy.}
\end{abstracts}

\vspace*{.25cm}
\section{Introduction: Lattice Approach to Heavy Quark Systems}
Heavy quark systems play an important r\^ole in the study of the less
well-known elements of the CKM matrix, which are central to our
understanding of the origin of CP violation, and may also contain
information of physics beyond the Standard Model. Set against this
background, the lattice formulation of QCD provides a framework for
the calculation of hadron masses and weak matrix elements from first
principles. Since the lattice approach is intrinsically
non-perturbative, one may tackle the large theoretical uncertainties
due to the strong interaction in weak decay amplitudes.


Lattice QCD replaces space-time by a four-dimensional, euclidean, hypercubic
lattice of size $L^3\cdot T$. The sites are separated by the lattice
spacing $a$, which acts as an UV cut-off. One problem encountered in
current simulations is that typical values of $a^{-1}$ lie in the range
2--$3.5\,\gev$. Therefore one expects that discretisation errors
(``lattice artefacts") will distort the results already for charm
physics. Also, $b$ quarks cannot be studied directly, since their mass
is above the UV cut-off.

Several methods are being used to circumvent this problem. It is now
customary for simulations in heavy quark physics to cancel the leading
discretisation error by employing so-called {\it improved\/} actions
and operators\,\cite{Improvement}, or by absorbing it into a rescaling
of the quark fields\,\cite{KLM_norm}. Quantities can then be computed
around $m_{{\rm charm}}$ and extrapolated to the $b$~quark mass.
Alternatively, one can use the {\it static approximation\/} and
perform the simulation at infinitely heavy quark mass, using the
leading term of the expansion of the heavy quark propagator in
$1/m_Q$.  Finally, one can employ a {\it non-relativistic} formulation
of the QCD Lagrangian. It is obvious that none of these methods is
entirely satisfactory, but that they provide complementary information
about heavy quark systems.

Apart from lattice artefacts, the main systematic errors in lattice
simulations include the effects due to neglecting internal quark loops
by using the so-called {\it quenched approximation}. The normalisation
of lattice operators is another source of systematic uncertainties;
due to the explicit breaking of chiral symmetry by the regularisation
procedure, lattice operators are in general related to their continuum
counterparts via (finite) normalisation constants, whose numerical
values are usually not known very precisely.  Also, the explicit
breaking of chiral symmetry and Lorentz invariance leads to mixing
with higher dimension operators. Hence, the connection between the
matrix element of a continuum operator $\widehat{O}$ and its lattice
counterparts is in general given by
\be
   \langle\widehat{O}\rangle^{\rm cont} = 
   \sum_i\,Z_i(g^2)\langle\widehat{O}_i\rangle^{\rm latt}
  + {\cal O}(a),
\ee
where the $Z_i$'s are the appropriate normalisation and matching factors.

Finally, lattice estimates of dimensionful quantities are subject to
uncertainties in the lattice scale. They arise from the fact that
different quantities which are used to set the scale $a^{-1}\,[\gev]$
give different results. This is closely related to using the quenched
approximation, since loop effects are not expected to be the same for
different quantities.

\section{Leptonic $B$ Decays and $\bbar$ Mixing}

The decay constants of heavy-light pseudoscalar and vector mesons,
$f_P$ and $f_V$, are related to the matrix elements of the lattice
axial and vector currents trough
\be
  \langle0|A_4^{\rm latt}(0)|P\rangle \sim M_P\,f_P/Z_A,\qquad
  \langle0|V_j^{\rm latt}(0)|V\rangle \sim 
                                      \epsilon_j\,M_V^2(f_V Z_V)^{-1},
\ee
where $Z_A,\,Z_V$ are the normalisation constants of the currents.
The matrix elements as well as the pseudoscalar and vector masses
$M_P,\,M_V$ are extracted from mesonic two-point correlation functions
through a fitting procedure.  HQET predicts scaling laws for the
combination $f_P\sqrt{M_P}$, which is expected to behave like a
constant as $M_P\rightarrow\infty$.  Furthermore, in the infinite mass
limit, the HQ-spin symmetry predicts that pseudoscalar and vector
decay constants become degenerate (up to short-distance corrections).
Hence
\be
  \widetilde{U}(M) \equiv \frac{f_Vf_P}{M}\Big/
  \left\{1+\textstyle\frac{2}{3\pi}\alpha_s(M)\right\} = 1+O(1/M)
\ee
It has been shown that $\widetilde{U}(M)$ indeed approaches one in the
infinite mass limit\,\cite{quenched}.  However, lattice studies have
revealed that the $1/M$ corrections to the scaling laws are large; for
the decay constant they amount to about 15\% at the $B$~mass and
about 40\% at the mass of the $D$~meson.

In Table\,1 we show the results for the decay constants
$f_D,\,f_{D_s}$ and $f_B$ from various lattice calculations using
relativistic heavy quarks.\footnote{We use a convention in which
  $f_\pi=132\,\mev$.} Expressing the results for the decay constant at
the individual values of~$a$ in terms of a common hadronic scale
$r_0\simeq0.5\,{\fm}$\,\cite{sommer}, one can extrapolate the
dimensionless quantity $f_P r_0$ to the continuum limit,
$a\rightarrow0$. Repeating the procedure using lattice data for
$f_\pi$, one obtains the continuum result from
\be
   f_P\,[\mev] = 132\,\mev\,\Big(f_P r_0/f_\pi r_0\Big)\Big|_{a=0}.
\ee
The results from this analysis using relativistic heavy quarks read
\bea   \label{eq:extra_fB}
  f_D &=& 205\pm45\,\mev,\qquad f_{D_s} = 220\pm50\,\mev  \nonumber\\
  f_B &=& 170\err{55}{50}\,\mev,\qquad f_{B_s}/f_B = 1.13\pm0.14.
\eea
\begin{table}[tb]  \label{tab:fB}
\caption{\small 
  Lattice estimates for the pseudoscalar decay constants from different
  collaborations. All data were obtained in the quenched approximation
  using relativistic heavy quarks.}
%
\begin{center}
\begin{tabular}{|c|c|llll|}
\hline
\hline
Collab. & $a\,[{\fm}]$ & $f_D\,[\mev]$ & $f_{D_s}\,[\mev]$ & $f_B\,[\mev]$
        & $f_{B_s}/f_B$ \\
\hline
\hline
MILC\,\cite{MILC_lat95} & 0
  &  182(3)(9)(22) & 198(5)(10)(19) & 151(5)(16)(26) & 1.11(2)(4)(8)
  \\
\hline
BLS\,\cite{BLS_93} & 0.059
  & 208(9)(35)(12) & 230(5)(10)(19) & 187(10)(34)(15) & 1.11(6) \\
\hline
LANL\,\cite{BG_95} & 0.083
  & 229(7)\err{20}{16} & 260(4)\err{27}{22} & & \\
& 0
  & 186(29)  & 218(15) & & \\
\hline
UKQCD\,\cite{quenched} & 0.069
  & 185\er{4}{3}\err{42}{7} & 212(4)\err{46}{7} & 160(6)\err{59}{19}
  & 1.22\er{4}{3} \\
 & 0.083
  & 199(15)\err{27}{19} & 225(15)\err{30}{22} & 176(25)\err{33}{15} 
  & 1.17(12) \\
\hline
ELC\,\cite{abada_92} & 0.051
  & 210(40) & 230(50) & 205(40) & \\
\hline
JLQCD\,\cite{JLQCD_lat95} & 0.059
  & 216(17) & 240(17) & 182(16) & \\
 & 0.078
  & 206(12) & 237(14) & 192(11) & \\
\hline
PCW\,\cite{PCW_prop_93} & 0
  & 170(30) & & 180(50) & 1.09(2)(5) \\
PCW\,\cite{PCW_91} & 0.083
  & 198(17) & 209(18) & & \\
\hline
\hline
\end{tabular}
\end{center}
\end{table}
The $B$ meson decay constant together with the $B$ parameter $B_B$ is
of great importance for the study of $\bbar$ mixing. The
renormalisation group invariant $B$ parameter $B_B$ is defined via
$
B_B = \alpha_s(\mu)^{-2/\beta_0}\,
        {\langle\overline{B^0}\left|\,O_L(\mu)\,\right|B^0\rangle}/
                  {\frac{8}{3}\,f_B^2\,M_B^2},
$
where $O_L(\mu)$ is the $\Delta B=2$ four fermion operator. In a recent
study, estimates for the $B$ parameter using the static approximation
were obtained\,\cite{bbar}
\be
   B_{B_d}=1.02\er{5}{6}\er{3}{2},\qquad
   B_{B_s}=1.04\er{4}{5}\er{2}{1},
\ee
where the first error is statistical, and the second is an estimate of
systematic errors. The authors attribute a further systematic error
of 15--20\% to the uncertainty in the perturbative matching factors.
The above result indicates that SU(3)-flavour breaking effects are small
for $B_B$, whereas they can be quite sizeable in the case of $f_B$.

These results can be applied to the ratio of $\bbar$ mixing parameters
$x_s/x_d$
\be
     \frac{x_s}{x_d} =
     \frac{\tau_{B_s}}{\tau_{B_d}}\,
     \frac{\hat\eta_{B_s}}{\hat\eta_{B_d}}\,\frac{M_{B_s}}{M_{B_d}}\,
     \frac{f_{B_s}^2\,B_{B_s}}{f_{B_d}^2\,B_{B_d}}\,
     \frac{|V_{ts}|^2}{|V_{td}|^2} =
     \big( 1.37 \pm 0.39 \big)\,
     \frac{|V_{ts}|^2}{|V_{td}|^2}.
\ee
In conjunction with the experimental value
$x_d=0.71(6)$\,\cite{PDG94}, the above results can also be used to
predict $x_s$, provided $\frac{|V_{ts}|^2}{|V_{td}|^2}$ is constrained
using global fits\,\cite{alilon}. Choosing $B_K=0.8\pm0.2$, and taking
our lattice estimate $f_B=170\pm55\,\mev$ one obtains
\be
   x_s = 13.4\pm4.0\errrr{10.3}{3.7}\errr{1.3}{0.7},
\ee
where the first error mainly reflects the error in the ratio
$f_{B_s}/f_B$, the second is due to the uncertainty in the actual
value of $f_B$, and the third arises from the uncertainty in
$B_K$. Clearly, much more precise values of $f_B$ are needed
in order to predict $x_s$ more reliably.

\newpage
\section{Semi-leptonic $B$ decays: heavy-to-light and heavy-to-heavy
  transitions }
Recently, there has been much activity in studying the decays
${B\rightarrow{\pi}\ell\overline{\nu}_\ell,\,{\rho}\ell\overline{\nu}_\ell}$,
which can be used to extract $V_{ub}$. The matrix elements for these
decays are parametrised
in terms of form factors, e.g. for \Bto{\pi}
\be
  \langle\pi|V_\mu|B\rangle = \big(p_B+p_\pi\big)_\mu\,f_+(q^2)
                             -\big(p_B-p_\pi\big)_\mu\,f_-(q^2),
\ee
where $q=p_{B}-p_{\pi}$ is the momentum transfer. For the decay
\Bto{\rho}, one has additional form factors $V,\,A_1,\,A_2$ and $A_3$.
An important ingredient in the analysis of these decays is the
observation that for infinite heavy quark mass, HQET predicts scaling
laws for the form factors near maximum momentum transfer $q^2_{\rm
  max}$, i.e. at leading order in $1/M$\,\cite{Isgur_Wise}
\be  \label{eq:HQff}
  f^+(q^2_{\rm max}) \sim M^{1/2},\quad
  V(q^2_{\rm max}) \sim M^{1/2},\quad
  A_1(q^2_{\rm max}) \sim M^{-1/2},\ldots
\ee
Due to the limitations imposed by the lattice spacing, simulations are
not yet suited for a direct computation of form factors for
${B\rightarrow{\pi}\ell\overline{\nu}_\ell,\,{\rho}\ell\overline{\nu}_\ell}$.
 One rather obtains the form factors for typical
lattice momenta $|\vec{p}|\le1.5\,\gev/c$ and for heavy quark masses
in the region of charm. Hence, the ``generic" semi-leptonic
heavy-to-light transition in current lattice simulations is
$D\rightarrow K\ell\overline{\nu}_\ell$, and typical momentum
transfers lie in the range $-0.8\,\gev^2/c^2\le
q^2\le1.7\,\gev^2/c^2$. Lattice results for form factors relevant for
semi-leptonic $D$~decays can be found in a recent 
review\,\cite{jns_lat95}.
\begin{table}[tp]  \label{tab:Btorhopi}

\caption{\small Lattice results for form factors relevant for
  semi-leptonic
  ${B\rightarrow\pi\ell\overline{\nu}_\ell,\,\rho\ell\overline{\nu}_\ell}$
  decays. All results are obtained using propagating heavy quarks with
  the leading discretisation errors subtracted, except for ELC.}
\begin{center}
\begin{tabular}{|c|c|r@{.}lr@{.}lr@{.}lr@{.}l|}
\hline
\hline
Collab. & $a$\,[fm] & \multicolumn{2}{c}{$f_+(0)$}
 & \multicolumn{2}{c}{$V(0)$} & \multicolumn{2}{c}{$A_1(0)$}
 & \multicolumn{2}{c|}{$A_2(0)$} \\
\hline
\hline
ELC\,\cite{abada_93} & 0.051
   & 0&30(14)(5) & 0&37(11) & 0&22(5) & 0&49(21)(5)  \\
\hline
APE\,\cite{APE_semi_94} & 0.083
   & 0&35(8) & 0&53(31) & 0&24(12) & 0&27(80) \\
\hline
UKQCD\,\cite{B_to_pi_95,B_to_Rho_95} & 0.069
 & 0&23(2) & \multicolumn{2}{c}{} 
 & 0&27\er{7}{4}\er{3}{3} & 0&28\er{9}{6}\er{4}{5} \\
\hline
GSS\,\cite{wupp_semi_95} & 0.059
 & 0&50(14)\er{7}{5} & 0&61(23)\er{9}{6} & 0&16(4)\err{22}{16}
 & 0&72(35)\err{10}{7} 
  \\
\hline
\hline
\end{tabular}
\end{center}
\end{table}
\begin{table}  \label{tab:slope_IW}
\caption{\small Lattice results for the slope parameter $\rho^2$ of the
  Isgur-Wise function, the applied parametrisation and, where quoted,
  estimates for $|V_{cb}|$ using measured decay rates. In the last
  column, we also list the form factor on which the estimate is
  based, or which method was used to formulate heavy quarks.}
\begin{center}
\begin{tabular}{|c|c|llcl|l|}
\hline
\hline
Collab. & $a$\,[fm] & $\rho_{u,d}^2$ & $\rho_s^2$ & Par. & $|V_{cb}|$ 
        &  \\
\hline
\hline
BSS\,\cite{BSS_93} & 0.059 &    & 1.21(26)(33) & lin. &
  0.044(5)(7) & $h_+$ \\
\hline
UKQCD\,\cite{iw_prd} & 0.069 & 0.9\er{2}{3}\er{4}{2} &
  1.2\er{2}{2}\er{2}{1} & BSW & 0.037(1)(2)\er{4}{1}  &
  $h_+$ \\
\hline
UKQCD\,\cite{hh+nmh_lat93} & 0.069 & 1.1(5) & 1.2\er{2}{3} & BSW &
  0.037(3)(5) & $h_{A_1}$ \\
\hline
LANL\,\cite{BG_semi_lat95} & 0.083 & 0.97(6) & & BSW & & $h_+$ \\
\hline
MO\,\cite{mandula_lat93} & 0.170 & & 0.95 & quad. & & static \\
\hline
Ken\,\cite{Ken_lat95} & 0.083 & 0.41(2) & & & &  static \\
\hline
HM\,\cite{HM_95} & 0.083 & 0.70(17) & & lin. & & NRQCD \\
\hline
\hline
\end{tabular}
\end{center}
\end{table}

In order to make predictions for
$B\rightarrow\pi\ell\overline{\nu}_\ell,\,\rho\ell\overline{\nu}_\ell$,
one needs to extrapolate the lattice form factors to the mass of the
$b$~quark using the above scaling laws.  Since the range of accessible
lattice momenta is rather restricted, i.e. $|\vec{p}|\ll m_b$, one
obtains the form factors in a narrow range of $q^2$ near $q^2_{\rm
  max}$.  Therefore, in order to determine the $q^2$-behaviour of form
factors for
$B\rightarrow\pi\ell\overline{\nu}_\ell,\,\rho\ell\overline{\nu}_\ell$,
or their values at $q^2=0$, one cannot avoid introducing a certain
model dependence in the lattice results: assuming vector pole
dominance is not reliable, since the accessible range of $q^2$ is
rather narrow, and lattice data can presently not distinguish between
different types of pole behaviour. In an alternative procedure, one
first interpolates the lattice data to $q^2=0$ for quark masses around
$m_{\rm charm}$. But in order to extrapolate the resulting form
factors at $q^2$ to the mass of the $b$~quark, one needs to guess its
leading scaling behaviour in the heavy mass, which, in contrast to
eq.\,(\ref{eq:HQff}), cannot be obtained from HQET. Several methods
have been applied, and Table\,2 lists the results for form factors at
$q^2=0$ from various groups.  It has been noted that the model
dependence could be avoided in the framework of light-cone sum
rules\,\cite{lightcone}, where one obtains a leading scaling behaviour
of $F(0)\sim M^{-3/2}$ for all form factors. This argument has so far
not been directly applied in lattice simulations.

It has also been suggested\cite{B_to_Rho_95} to use lattice form
factors to calculate the differential decay rate for the 
{\it exclusive\/} decay
$\overline{B}^0\rightarrow\rho^+\ell^-\overline{\nu}_\ell$ beyond the
region of charm production. One can then avoid the difficult
determination of $F(q^2=0)$, and a model independent extraction of
$|V_{ub}|$ is possible using experimental data for the differential
decay rate $d\Gamma/dq^2$
%
\be
|V_{ub}|^{-2}\frac{d\Gamma}{dq^2}  \propto 
  \left\{|H^+(q^2)|^2+|H^-(q^2)|^2+|H^0(q^2)|^2\right\} 
  = 
  {\cal A}^2\left(1+{\cal B}(q^2-q^2_{\rm max})\right),
\ee
where 
$H^\pm(q^2)$ and $H^0(q^2)$ are combinations of the form factors
$A_1(q^2),\,A_2(q^2)$ and $V(q^2)$. The term 
${\cal A}^2(1+{\cal B} (q^2-q^2_{\rm max}))$ parametrises
long-distance hadronic dynamics. In fact,
${\cal A}^2$ plays the same r\^ole as the Isgur-Wise function in
the case of heavy-to-heavy transitions. Here, lattice data can be used
to determine the normalisation, whereas in the heavy-to-heavy case,
the overall normalisation is provided by the HQ-symmetry. In
a recent study\cite{B_to_Rho_95}, the authors obtain 
${\cal A}^2=21\pm3\;\gev^2$ and 
${\cal B}=(-8\er{4}{6})\,10^{-2}\;\gev^{-2}$.

Heavy-to-heavy transitions like
$B\rightarrow{D}\ell\overline{\nu}_\ell,\,{D^*}\ell\overline{\nu}_\ell$
are also parametrised in terms of six form factors, $h_+, h_-, h_V,
h_{A_1}, h_{A_2}$ and $h_{A_3}$, which are functions of $\w$, the
product of 4-velocities of the $B$ and $D$ mesons. HQ-symmetry relates
the six form factors to one universal form factor, $\xi(\w)$, called
the Isgur-Wise function, which is normalised at zero recoil,
$\xi(1)=1$. The form factor $h_+(\w)$ is related to $\xi(\w)$ via
\be
    h_+(\w) = (1+\beta_+(\w)+\gamma_+(\w))\,\xi(\w),
\ee
where $\beta_+(\w)$ parametrises radiative corrections between HQET and full
QCD, and $\gamma_+(\w)$ denotes (unknown) corrections in the inverse
heavy quark mass.

On the lattice one typically obtains the form factors from a ratio of
the relevant matrix elements at $\w$ and at zero recoil ($\w=1$).
Thereby, some of the systematic effects cancel, most notably the
normalisation of the axial and vector currents.  In order to obtain an
estimate for $\xi(\w)$, the known radiative corrections can also be
subtracted. Lattice data for the form factors $h_+$ and $h_{A_1}$ have
been used to test the HQ-symmetry \cite{lpl_lat94,iw_prd}. It has been
found that $h_+(\w)$ shows only a weak dependence on the heavy quark
mass, and that the Isgur-Wise function extracted from both $h_+$ and
$h_{A_1}$ is compatible within statistical errors. This may be
interpreted as a manifestation of the HQ-spin-flavour symmetry.

In order to extract $|V_{cb}|$ from the experimentally measured decay
rate for \Bto{D^*} using lattice data, one first needs to parametrise
$\xi(\w)$. One particular parametrisation is
\be
    \xi_{\rm BSW}(\w) = \textstyle\frac{2}{\w+1}\,
       \exp\left\{ -(2\rho^2-1)\textstyle\frac{\w-1}{\w+1} \right\},
\ee
where $\rho^2$ is the slope of $\xi(\w)$ at zero recoil. Other
parametrisations, based on linear, quadratic and pole {\it ans\"atze},
have also been studied. Lattice data have so far not revealed any significant
dependence on the chosen parametrisation\cite{jns_lat93,iw_prd}.

Table\,3 contains lattice results for the slope, either for massless
spectator quarks ($\rho^2_{u,d}$), or in the case that the spectator
quark is a strange quark ($\rho^2_s$). Also, the parametrisation and
the estimates of $|V_{cb}|$ are given. Lattice results for the slope
are consistent among the different collaborations, although the errors
are still large.  Recently, preliminary results\,\cite{peer_eps95} for
semi-leptonic $\Lambda_b\rightarrow\Lambda_c$ decays have been shown,
which will be valuable in further studies of the HQ-symmetry.

\begin{table}    \label{tab:UKQCD_baryons}
\caption{\small Masses and mass splittings of heavy baryons from
  various collaborations.}
\vspace{0.2cm}
\begin{center}
\begin{tabular}{|c|c|r@{.}lr@{.}l|r@{.}lr@{.}l|}
\hline
\hline
 &  & \multicolumn{4}{c|}{$h=$charm}
    & \multicolumn{4}{c|}{$h=$beauty} \\
 & Collab. & \multicolumn{2}{c}{Latt.\,[\gev]}
                 & \multicolumn{2}{c|}{Exp.\,[\gev]}
                 & \multicolumn{2}{c}{Latt.\,[\gev]}
                 & \multicolumn{2}{c|}{Exp.\,[\gev]} \\
\hline
\hline
$\Lambda_{h}$ & UKQCD\,\cite{baryons}
  & 2&27\er{4}{3}\er{3}{3} & 2&285(1)
  & 5&64\er{5}{5}\er{3}{2} & 5&641(50) \\
              & PCW\,\cite{PCW_baryons}
  &  \multicolumn{2}{c}{} &  \multicolumn{2}{c|}{}
  & 5&728$\pm$0.144$\pm$0.018 &  \multicolumn{2}{c|}{} \\
\hline
$\Sigma_{h}$  & UKQCD\,\cite{baryons}
  & 2&46\er{7}{3}\er{5}{5} & 2&453(1) 
  & 5&77\er{6}{6}\er{4}{4} & 5&814(60) \\
$\Sigma^*_h$  & UKQCD\,\cite{baryons}
  & 2&44\er{6}{4}\er{4}{5} & 2&530(7)
  & 5&78\er{5}{6}\er{4}{3} & 5&870(60) \\
$\Xi_{h}$     & UKQCD\,\cite{baryons}
  & 2&41\er{3}{3}\er{4}{4} & 2&468(4)
  & 5&76\er{3}{5}\er{4}{3} & \multicolumn{2}{c|}{} \\
$\Omega_{h}$  & UKQCD\,\cite{baryons}
  & 2&68\er{5}{4}\er{5}{6} & 2&704(20)
  & 5&99\er{5}{5}\er{5}{5} & \multicolumn{2}{c|}{} \\
\hline
\hline
 & Collab. & \multicolumn{2}{c}{Latt.\,[\mev]}
           & \multicolumn{2}{c|}{Exp.\,[\mev]}
           & \multicolumn{2}{c}{Latt.\,[\mev]}
           & \multicolumn{2}{c|}{Exp.\,[\mev]} \\ 
\hline
$\Lambda_h-P$ & UKQCD\,\cite{baryons} 
              & \multicolumn{2}{l}{408\err{41}{31}\err{34}{35}}
              & \multicolumn{2}{c|}{417(1)}
              & \multicolumn{2}{l}{359\err{55}{45}\err{27}{26}}
              & \multicolumn{2}{c|}{362(50)} \\
              & ELC\,\cite{boch_91} 
              & \multicolumn{2}{l}{} 
              & \multicolumn{2}{l|}{} 
              & \multicolumn{2}{l}{$720\pm160$\errr{0}{130}}
              & \multicolumn{2}{l|}{}  \\
              & UKQCD\,\cite{bbar} 
              & \multicolumn{2}{l}{} 
              & \multicolumn{2}{l|}{} 
              & \multicolumn{2}{l}{420\errr{100}{90}\err{30}{30}}
              & \multicolumn{2}{l|}{} \\
              & PCW\,\cite{PCW_baryons} 
              & \multicolumn{2}{l}{$564\pm88\pm18$}
              & \multicolumn{2}{l|}{} 
              & \multicolumn{2}{l}{$458\pm144\pm18$}
              & \multicolumn{2}{l|}{} \\
\hline
$\Sigma_h-\Lambda_h$ & UKQCD\,\cite{baryons} 
              & \multicolumn{2}{l}{190\err{50}{43}\err{13}{13}}
              & \multicolumn{2}{c|}{169(2)}
              & \multicolumn{2}{l}{157\err{52}{64}\err{11}{11}}
              & \multicolumn{2}{c|}{173(11)} \\
\hline
$\Sigma_h^*-\Sigma_h$  & UKQCD\,\cite{baryons} 
              & \multicolumn{2}{l}{--17\err{12}{31}\er{3}{2}}
              & \multicolumn{2}{c|}{77(6)}
              & \multicolumn{2}{l}{--6\err{4}{11}\er{1}{1}}
              & \multicolumn{2}{c|}{56(16)} \\ 
$\Xi_h^*-\Xi_h^\prime$ & 
              & \multicolumn{2}{l}{--20\err{12}{24}\er{2}{3}}
              & \multicolumn{2}{c|}{83}
              & \multicolumn{2}{l}{--7\er{4}{8}\er{1}{1}}
              & \multicolumn{2}{c|}{}   \\
$\Omega_h^*-\Omega_h^\prime$ & 
              & \multicolumn{2}{l}{--23\err{6}{14}\er{3}{2}}
              & \multicolumn{2}{c|}{}
              & \multicolumn{2}{l}{--8\er{2}{5}\er{1}{1}}
              & \multicolumn{2}{c|}{}  \\
\hline
\hline
\end{tabular}
\end{center}
\end{table}

\section{Heavy Baryon Spectroscopy}
There has been increased experimental and theoretical activity in the
study of baryons containing one heavy quark. Lattice calculations can
make predictions for the masses of heavy baryons, which are extracted
from the exponential fall-off of correlation functions of suitably
chosen interpolating operators. The results from two recent studies
\cite{PCW_baryons,baryons} are listed in Table\,4, and one observes
good agreement with experimental data. Mass splittings involving heavy
baryons have also been studied by a number of groups, and the results
are also given in Table\,4. The agreement between experimental and
lattice data for the $\Lambda-$ Pseudoscalar and $\Sigma-\Lambda$
splittings is rather good. However, the spin splittings, in particular
for $\Sigma_c^*-\Sigma_c$, are definitely inconsistent with experiment.
This is attributed to a convolution of lattice artefacts and the
quenched approximation.

\vspace*{0.5cm}
\par\noindent
{\bf 5. Acknowledgements}
\par
I would like to thank the organisers of this workshop for creating
such a pleasant and creative atmosphere. Support by the
Heisenberg-Landau Programme is gratefully acknowledged.

\vspace*{0.5cm}
\par\noindent
{\bf 6. References}

\end{document}